\documentclass{article}
\usepackage{ijcai26}

\usepackage{times}
\usepackage{soul}
\usepackage{url}
\usepackage[hidelinks]{hyperref}
\usepackage[utf8]{inputenc}
\usepackage[small]{caption}
\usepackage{graphicx}
\usepackage{amsmath}
\usepackage{amsthm}
\usepackage{amssymb}
\usepackage{booktabs}
\usepackage{algorithm}
\usepackage{algorithmic}
\usepackage[switch]{lineno}
\usepackage{makecell}
\usepackage{amsfonts}
\usepackage{stfloats}

\title{Learning Collective Medication Effects via Multi-level Abstraction for Medication Recommendation}

\author{
Yanda Wang$^1$
\and
Weitong Chen$^2$\and
Chao Tan$^1$\and
Ian Nabney$^3$\and
Lin Yue$^2$\and
Genlin Ji$^1$
\\
\affiliations
$^1$Nanjing Normal University\\
$^2$Adelaide University\\
$^3$University of Bristol\\
\emails
yandawang@nnu.edu.cn,
\{weitong.chen,lin.yue\}@adelaide.edu.au,
ian.nabney@bristol.ac.uk
}

\begin{document}

\maketitle

\begin{abstract} 


Historical prescriptions and selected candidate drugs relevant to the current visit serve as important references for medication recommendation. However, in the absence of explicit intrinsic principles for semantic composition, existing methods treat synergistic drugs as independent entities and fail to capture their collective therapeutic effects, resulting in a mismatch between medication-level references and longitudinal patient representations. In this paper, we propose MSAM, a novel medication recommendation model that bridges the gap via multi-level medication abstraction. The model introduces a multi-head graph reasoning mechanism to organize flat daily medication sets into clinically meaningful semantic units, serving as intermediate abstraction results to propagate features from individual drugs to higher-level representations. Building on these units, MSAM performs two-stage abstraction over historical prescriptions and selected candidates via intra- and inter-level feature propagation across heterogeneous clinical structures, capturing collective therapeutic effects aligned with patient conditions. Experiments on two real-world clinical datasets show that MSAM consistently outperforms state-of-the-art methods, validating the effectiveness of structural medication abstraction for recommendation.




\end{abstract}

\section{Introduction}

Recent advances in deep learning have driven substantial progress across different healthcare applications \cite{saiful2025accessible,ye2025medualtime,qu2025multimodal}, including medication recommendation that generates effective prescriptions from patient histories \cite{wang2025beyond,liang2025cidgmed,li2024stratmed}. Clinical practice shows that integrating patient conditions with clinically relevant medications, which are drawn from historical prescriptions and candidate drugs, can improve recommendation performance, since these medications reflect treatment strategies for similar conditions \cite{kim2024vita,wang2021self}. Nevertheless, aligning patient representations encoding rich health conditions with flat medication sets remains a fundamental challenge due to the clinical semantic gap between them.

Owing to widespread multimorbidity in clinical practice, where patients suffer different co-existing diseases, patient conditions are summarized from multiple clinical events into high-level representations that capture complex health states \cite{yu2024aka,mi2024acdnet}. 
Meanwhile, medications are typically prescribed simultaneously and work in combination to implement coordinated treatment strategies, whose effects arise from interactions among individual drugs when organized into structured representations. Such collective effects manifest at multiple granularities, ranging from single drugs, to daily co-prescribed sets, and further to prescription-level patterns across visits. 
However, while these structures are implicit in clinical practice, unlike domains such as natural language, medications lack intrinsic principles for clinical semantic composition. Existing methods have to treat co-prescribed drugs as independent labels rather than organizing them into meaningful therapeutic representations, failing to capture their collective effects \cite{wang2024topological,kim2024vita,shang2019gamenet}. Consequently, existing methods use high-level patient representations to retrieve individual drugs and then integrate medication-level effects as references for complex patient conditions. This mismatch between individually referenced medications and structured patient representations cannot be resolved by direct patient–drug matching alone.

To bridge the semantic gap, we propose \textbf{M}ulti-level \textbf{S}tructure-aware \textbf{A}bstraction of \textbf{M}edication effects (\textbf{MSAM}), a novel recommendation model that structurally abstracts medications into references aligned with high-level patient representations. The key insight originates from the inherent multi-level organization of clinical practice: individual medications have distinct effects, sets of medications collectively address specific conditions and form daily prescriptions, and prescriptions spanning multiple days yield visit-level effects. The central challenge arises at the daily level: while the assignment of medications to specific days and visits is clinically defined, medications prescribed within the same day form an unordered set without intrinsic semantic structures, rendering feature propagation from individual drugs to higher-level abstractions ill-defined. In addition, abstraction must operate under two constraints: the process involves heterogeneous clinical relationships, including graph-based interactions at the medication level, temporal dependencies within and across days, and relevance of historical prescriptions to the current visit; while historical prescriptions provide observed medications for abstraction, abstraction over the broad set of candidates requires prior identification of relevant medications, leading to a dependency that prevents joint optimization of identification and abstraction.

To address these issues, MSAM introduces a Graph Attention Networks (GAT)-based multi-head graph reasoning mechanism that organizes flat daily medication sets into clinically meaningful intermediate semantic units, enabling well-defined propagation of medication features from individual drugs to higher-level abstractions. 
Building on these units, MSAM performs multi-level medication abstraction to propagate their features across heterogeneous structures, so that it captures prescription therapeutic effects and allows visit conditions to query semantically aligned prescription-level treatments. To address the dependency that candidates must be identified before abstraction, MSAM performs the abstraction in two stages: first, abstract historical prescriptions and identify candidates, followed by abstraction of selected medications. Experiments on two real-world clinical datasets show that MSAM consistently outperforms competitive baselines, remains robust across varying sizes of medication sets, and improves performance of existing recommendation models when integrated with them, indicating its ability to model and align collective medication effects with patient conditions for improved recommendation.

The contributions of this work can be summarized as follows:

\begin{itemize}
    \item We highlight that explicitly abstracting inherent multi-level structures within medications exhibits their collective therapeutic effects, which align with the meticulously encoded patient representations to act as informative references for medication recommendation.
    \item We propose MSAM, a novel two-stage medication recommendation framework that performs multi-level structural abstraction of medications despite the absence of explicit semantic rules to bridge the clinical semantic gap between references and patient conditions.
    \item We conduct extensive experiments on two widely used real-world clinical datasets, the results indicate that MSAM consistently surpasses baseline methods in terms of different metrics.
\end{itemize}

\section{Related Work}


\subsection{Medication Recommendation}

Early studies on medication recommendation primarily rely on Electronic Healthcare Records (EHRs) from a single hospitalization visit to generate prescriptions, commonly referred to as instance-based methods \cite{zhang2017leap}. Focusing on the patient conditions within one visit, these methods ignore condition progressions across multiple visits and often yield suboptimal performance. 
In contrast, longitudinal-based methods exploit patient histories spanning multiple visits to capture inherent temporal dependencies, therefore summarizing condition trajectories to perform more accurate recommendation \cite{yang2025patient,yu2024aka,mi2024acdnet}.
In addition to EHRs-centric methods, another line of research incorporates drug structures to enhance medication representation \cite{mu2025medication,zou2024dainet}. By encoding molecular structures and drug–drug interactions, these methods uncover pharmacological properties of medications to improve accuracy and safety.

\subsection{Medication Reference for Recommendation}

Recent studies indicate that incorporating medications clinically relevant to the current patient condition as references can further improve recommendation performance \cite{kim2024vita,shang2019gamenet}, and existing methods generally consider two primary sources of references \cite{kim2024vita,shang2019gamenet,wang2021self}. One line of work retrieves medications from historical prescriptions based on the clinical relevance between the current visit and past visits, assuming that medications issued for similar conditions act as informative guidance. To ensure effective retrieval, various strategies have been proposed, such as multi-hop reading, coverage-based attention distribution over prescriptions, and adaptively determined reading hops \cite{wang2021self,wang2022adaptive}. The other line of work directly selects references from a broad set of candidate medications using the current patient representation as a query based on the attention mechanism, guiding the model to focus on drugs most related to the current patient condition \cite{wang2025beyond}.

\section{Proposed Method}



\begin{figure*}
    \centering
    \includegraphics[width=1\linewidth]{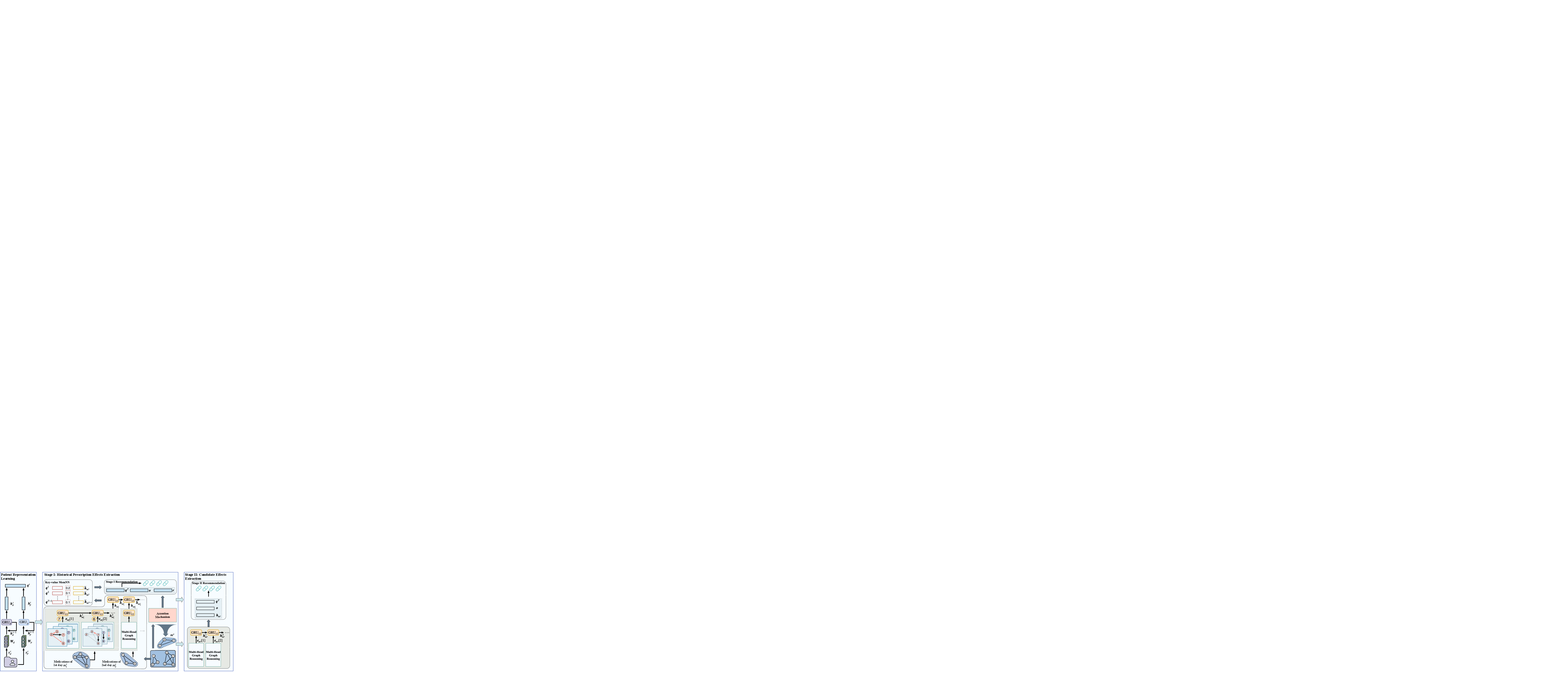}
    \caption{Workflow of MSAM}
    \label{fig:workflow}
\end{figure*}

In this section, we introduce details of MSAM, which employs a two-stage multi-level medication abstraction to extract collective effects from historical prescriptions and candidate medications, treating them as informative references aligned with patient conditions to perform the recommendations. An overview of MSAM is shown in Figure \ref{fig:workflow}.

\subsection{Notations}

\noindent\textbf{Patient Records:} EHRs describe each patient as a sequence of multivariate observations $X=[x^1, x^2,..., x^T]$, where $x^t$ represents records of the $t^{th}$ visit and $T$ is the number of total visits. $x^t$ consists of clinical events diagnoses $d^t$, procedures $p^t$ and medications $m^t$, which are the subsets of three vocabularies $S_d$, $S_p$ and $S_m$ respectively. Each vocabulary maps distinct clinical events into unique integers indices.

\noindent\textbf{EHRs Graph:} The EHRs graph $G_E=\{S_m, \mathcal{E}\}$ indicates co-occurrence relationships of medications across the whole patient cohort, where $\mathcal{E}$ denotes the set of edges between medications. The corresponding adjacency matrix is denoted as $A_E\in \mathbb{R}^{|S_m|\times|S_m|}$, where $A_E[i,j]=A_E[j,i]=1$ if the medication $i$ and $j$ have been prescribed together in a single visit, and 0 otherwise.

\noindent\textbf{Medication Recommendation:} Given the historical records $[x^1, x^2,..., x^{T-1}]$, the diagnoses $d^T$ and procedures $p^T$ of the current visit, and EHRs graph $G_E$, MSAM aims to predict medications $m^T$.

\subsection{Patient Representation Learning}
MSAM encodes a patient's historical records to build their representations for the recommendation. The model first transforms $d^t$ and $p^t$ into multi-hot vectors $\boldsymbol{s}_d^t\in \mathbb{R}^{|S_d|}$ and $\boldsymbol{s}_p^t\in \mathbb{R}^{|S_p|}$ respectively, which are then projected into continuous embeddings using matrices $\boldsymbol{W}_d\in \mathbb{R}^{|S_d|\times dim}$ and $\boldsymbol{W}_p\in \mathbb{R}^{|S_p|\times dim}$, where $dim$ is the embedding dimension, yielding event-level semantics. To abstract visit-level representations, MSAM aggregates the embeddings within each visit via average and feeds the results into GRU to model modality-specific temporal dependencies across visits.

\begin{equation}
    \boldsymbol{h}_d^t=\mathrm{GRU}_d(\boldsymbol{h}_d^{t-1}, \mathrm{Avg}(\boldsymbol{s}_d^t \boldsymbol{W}_d))
    \label{equ: encode diagnoses}
\end{equation}

\begin{equation}
    \boldsymbol{h}_p^t=\mathrm{GRU}_p(\boldsymbol{h}_p^{t-1}, \mathrm{Avg}(\boldsymbol{s}_p^t \boldsymbol{W}_p))
    \label{equ: encode procedures}
\end{equation}

\noindent MSAM finally integrates $\boldsymbol{h}_d^t$ and $\boldsymbol{h}_p^t$ with a linear layer to build visit-level representation $\boldsymbol{q}^t$. Therefore, event features propagate from event-level embeddings, modality-specific temporal modeling, to unified visit-level representations.

\subsection{Medication Effects Extraction and Reference}
\label{subsec: medication effects extraction and reference}

Unlike existing graph-based methods that model medication relationships to obtain individual drug embeddings, MSAM organizes flat sets of drugs via graph traversal to capture their collective therapeutic effects as intermediate semantic units, which enable feature propagation from individual medications to higher-level representations and yield references aligned with patient conditions. This is a two-stage framework: Stage I abstracts observed historical prescriptions and trains the attention-based candidate selection module, while Stage II selects and abstracts candidate medications.

\noindent\textbf{A. Stage I: Historical Prescription Effects Extraction}

When performing multi-level abstraction on historical prescriptions, MSAM treats individual drugs as initial nodes in Level 1 ($L1$) that represent their single therapeutic effects, and leverages EHRs graph reasoning to organize sets of medication nodes into sequences to uncover combination effects for specific conditions, while multiple consecutive sequences within a day are abstracted into the daily node in $L2$. Then daily nodes across different days of a visit are further aggregated into the prescription node in $L3$, and nodes of multiple historical prescriptions are summarized as the final reference node in $L4$ based on the clinical relevance between historical visits and current visit. Medication features propagate within and across multiple levels through graph structures, temporal dependencies, and clinical relevance.

\noindent\textbf{1) Medication-Level Semantics}

MSAM extracts medication features to capture their individual therapeutic effects by considering both global and local relationships. Specifically, the model embeds medications in $S_m$ via the matrix $\boldsymbol{W}_m\in \mathbb{R}^{|S_m| \times dim}$ and produces continuous representations $\boldsymbol{e}_m\in \mathbb{R}^{|S_m| \times dim}$, which are then updated using the GAT over EHRs graph $G_E$:

\begin{equation}
    \boldsymbol{e}_m=\mathrm{GAT}(\boldsymbol{e}_m, G_E)
    \label{equ: update medication embeddings over EHRs graph}
\end{equation}

\noindent Next, MSAM updates $\boldsymbol{e}_m$ based on local relationships within a prescription $m^t$ before abstracting it into the next level:

\begin{equation}
    \boldsymbol{e}_m=\mathrm{GAT}(\boldsymbol{e}_m, G_{E,m^t})
    \label{equ: update medication embeddings within a prescription}
\end{equation}

\noindent where $G_{E,m^t}$ is the subgraph corresponding to $m^t$. Through this process, medication features propagate along both global and local graph structures, and allow MSAM to capture their relationships across the entire patient cohort and within each prescription, generating embeddings based on clinical relevance to reveal individual drug effects.

\noindent\textbf{2) Set-Level Semantics via Multi-Head Graph Reasoning}

Without intrinsic structures to organize individual drugs into semantically meaningful representations, a flat set of medications $m$ cannot capture collective effects and limits feature propagation from individual drugs to higher-level abstraction. MSAM designs the GAT-based multi-head graph reasoning mechanism to traverse the EHRs graph and induce a ordered traversal over $m$, yielding representations that support semantic composition and cross-level feature propagation. Given the GAT parameters and EHRs graph, the mechanism aggregates diverse interaction patterns rather than relying on a single attention head to stabilize the ordering.

Specifically, the model first extracts the subgraph $G_{E,m}$ corresponding to $m$, and then starts the reasoning by selecting the initial medication $\mathcal{S}_{m,1}$ with the highest attention weight with respect to the query, whose formulation varies depending on the abstraction stage and is detailed in subsequent sections. For each subsequent drug $\mathcal{S}_{m,i}$ at step $i$, MSAM identifies its unvisited neighboring nodes $\mathcal{N}(\mathcal{S}_{m,i})$ in $G_{E,m}$, and applies the attention head $\overline{h}$ of GAT to compute the weights $\boldsymbol{\alpha}_{i,\overline{h}}$ between $\mathcal{S}_{m,i}$ and $\mathcal{N}(\mathcal{S}_{m,i})$, allowing it to vote for the drug with the highest weight. Since different heads capture diverse medication interaction patterns, aggregating votes across multiple heads enables the model to reach a consensus and yield the selection of  $\mathcal{S}_{m,i+1}$.

In case where multiple medications from $\mathcal{N}(\mathcal{S}_{m,i})$ receive the same maximum votes, let $\mathcal{H}_{i,\hat{i}}$ denote the set of attention heads that vote for $\mathcal{N}(\mathcal{S}_{m,i})[\hat{i}]$, MSAM averages its weights from these heads as shown in Eq.(\ref{equ: average attention from multiple heads}), and chooses the drug with the highest mean attention weights as $\mathcal{S}_{m,i+1}$.

\begin{equation}
    \hat{\boldsymbol{\alpha}}_{i,\hat{i}}=\frac{1}{|\mathcal{H}_{i,\hat{i}}|}\sum_{\overline{h}\in\mathcal{H}_{i,\hat{i}}}\boldsymbol{\alpha}_{i,\overline{h}}[\hat{i}]
    \label{equ: average attention from multiple heads}
\end{equation}

\noindent The iterative process continues all nodes in $G_{E,m}$ are visited and result in the ordered sequence $\mathcal{S}_m$.

MSAM applies the above graph reasoning to organize $m^t_i$, the set of medications prescribed on the $i^{th}$ day of the $t^{th}$ visit, into $\mathcal{S}_{m^t_i}$ to capture set-level semantics, and selects corresponding embeddings $e_{m^t_i}$ to updates them using GAT:

\begin{equation}
    \boldsymbol{e}_{m^t_i}=\mathrm{GAT}(\boldsymbol{e}_{m^t_i}, G_{E,m^t_i})
    \label{equ: update medication embeddings within a day}
\end{equation}

\noindent Subsequently, MSAM feeds these embeddings to GRU and encodes the organized sequence:

\begin{equation}
    \boldsymbol{h}_{m^t_i}^j=\mathrm{GRU}_{L2}(\boldsymbol{h}_{m^t_i}^{j-1}, \boldsymbol{e}_{m^t_i}[j])
    \label{equ: encode medications within a day}
\end{equation}

\noindent where $\boldsymbol{h}_{m^t_i}^j$ is the hidden state for the medication $\mathcal{S}_{m^t_i,j}$, and $\boldsymbol{e}_{m^t_i}[j]$ denotes its embedding. For simplicity, the final daily node for $m^t_i$ is denoted as $\boldsymbol{h}_{m^t_i}$. Through this process, medication features are propagated and abstracted into the daily prescription via sequential modeling.

Without explicit supervision for organizing flat medication sets, the GAT-based graph reasoning module is optimized in the end-to-end manner together with the overall model, capturing clinically meaningful medication relationships to organize medication sets into structured sequences.

\noindent\textbf{3) Multi-Level Prescription Abstraction}




Based on the medication-level and set-level semantics, MSAM further abstracts each historical prescription by encoding temporal dependencies between days, and finally distills clinically relevant effects across all prescriptions into a unified reference to complete the multi-level abstraction.

To abstract medications $m^t_1$ on the first day of $m^t$, MSAM uses $\boldsymbol{q}^t$ as the query and calculates attention weights $\boldsymbol{\alpha}_{m^t_1}$ to select the drug with highest weight as the starting node $\mathcal{S}_{m^t_1,1}$ for graph reasoning. 

\begin{equation}
    \boldsymbol{\alpha}_{m^t_1}=\mathrm{Softmax}(\boldsymbol{q}^t \boldsymbol{W}_{m} \boldsymbol{e}_{m^t_1})
    \label{equ: attention to select the first medication}
\end{equation}

\noindent The model then iteratively traverses the subgraph $G_{E,m^t_1}$ to organize $m^t_1$ into the structured sequence, from which the daily node $\boldsymbol{h}_{m^t_i}$ is generated. The attention module is shared between Stage I and Stage II: it selects the initial drug from $m^t_1$ and is later reused to select relevant candidate medications, ensuring a consistent selection across stages.

To abstract medications $m^t_i$ for subsequent days, MSAM treats the last medication from the sequence $\mathcal{S}_{m^t_{i-1}}$ of previous day as an additional node in the subgraph $G_{E,m^t_i}$ and uses it as the starting node for multi-head graph reasoning, organizing $m^t_i$ into sequence $\mathcal{S}_{m^t_{i}}$ to obtain the daily node $\boldsymbol{h}_{m^t_i}$. This design explicitly incorporates information from previous treatments when organizing current medications.

After obtaining daily nodes for all days of a prescription, MSAM feeds them into GRU to model temporal dependencies across multiple days within the visit:

\begin{equation}
    \overline{\boldsymbol{h}}_{m^t_i}=\mathrm{GRU}_{L3}(\overline{\boldsymbol{h}}_{m^t_{i-1}},\boldsymbol{h}_{m^t_i})
    \label{equ: encode medications acorss multiple days}
\end{equation}

\noindent For simplicity, MSAM denotes the last hidden state as $\overline{\boldsymbol{h}}_{m^t}$ to represent the abstraction result of prescription $m^t$.

Given all abstracted historical prescriptions, MSAM further builds a key-value Memory Neural Network (MemNN) $M$, whose keys $M_k$ are $\boldsymbol{q}^t (t\in [1, T-1])$ and values $M_v$ are corresponding abstracted prescriptions $\overline{\boldsymbol{h}}_{m^t}$, to extract relevant prescription effects. MSAM uses the current visit representation $\boldsymbol{q}^T$ as the query, and computes attention weights $\boldsymbol{\alpha}_M$ over the keys $M_k$.


\begin{equation}
    \boldsymbol{\alpha}_M=\mathrm{Softmax}((\boldsymbol{q}^T)^{\intercal} \boldsymbol{q}^{\ast})
    \label{equ: MemNN keys attention weight}
\end{equation}

\noindent $\boldsymbol{\alpha}_M$ is then used to identify relevant historical visits and abstract their prescription effects from $M_v$ into the reference $\boldsymbol{o}$.

\begin{equation}
    \boldsymbol{o}=\boldsymbol{\alpha}_M \overline{\boldsymbol{h}}_{m^{\ast}}
    \label{equ: read from MemNN values}
\end{equation}

Through the multi-level abstraction, MSAM propagates medication features from individual drugs to daily summaries, prescription-level and final reference, yielding comprehensive therapeutic effects aligned with patient conditions.

\noindent\textbf{4) Recommendation via Prescription Effects Reference}

In stage I, MSAM performs the recommendation by jointly referring to candidate medications and abstracted historical prescriptions. Note that candidate medications are referenced as medication-level embeddings, since multi-head graph reasoning is designed to extract collective effects from medications working in combination, and applying the abstraction over all candidates would introduce spurious interactions. Therefore, identifying medications relevant to the current visit is a prerequisite for candidate effect abstraction, which is deferred to Stage II.

Given the embedding $\boldsymbol{e}_m$ of all medications, MSAM uses $\boldsymbol{q}^T$ as the query to calculate attention weights over them, reusing the same attention module, i.e., $\boldsymbol{W}_m$, as in Eq.(\ref{equ: attention to select the first medication}) for selecting the first medication for prescription abstraction.

\begin{equation}
    \boldsymbol{\alpha}_m=\mathrm{Softmax}(\boldsymbol{q}^T \boldsymbol{W}_m \boldsymbol{e}_m)
    \label{equ: candidate medication attention}
\end{equation}

\noindent which are used to read relevant medications from $\boldsymbol{e}_m$.

\begin{equation}
    \boldsymbol{r}=\boldsymbol{\alpha}_m \boldsymbol{e}_m
    \label{equ: read from candidate medications}
\end{equation}

\noindent Then MSAM integrates patient representation $\boldsymbol{q}^T$, abstracted prescriptions $\boldsymbol{o}$, and read candidates $\boldsymbol{r}$ to perform the recommendation, where $\sigma(\cdot)$ is a sigmoid function, $f(\cdot)$ is the linear layer, and $\hat{\boldsymbol{y}}\in \mathbb{R}^{|S_m|}$ is the probability vector. Drugs with a probability exceeding 0.5 form the generated prescription $\hat{Y}$.

\begin{equation}
    \hat{\boldsymbol{y}}=\sigma(f([\boldsymbol{q}^T, \boldsymbol{o}, \boldsymbol{r}]))
    \label{equ: stage I recommendation}
\end{equation}

\noindent\textbf{B. Stage II: Candidate Effects Extraction}

In Stage II, MSAM extends abstraction to candidate medications. Based on the attention module trained in Stage I, the model first selects candidates relevant to the current visit, ensuring only those medications pertinent to the patient condition are considered, and then abstracts them to integrate with historical prescription references for recommendation.

\noindent\textbf{1) Candidate Selection and Abstraction}

Using $\boldsymbol{q}^T$ as the query, MSAM computes attention weights over all candidate medications based on the attention module trained in Eq.(\ref{equ: candidate medication attention}), and selects those exceeding the predefined threshold to form $m^s$. Next, the model abstracts $m^s$ by first obtaining their embeddings $\boldsymbol{e}_{m^s}$ to capture individual effects, and then applying the multi-head graph reasoning strategy introduced in Sec.\ref{subsec: medication effects extraction and reference}.A.2). This organizes $m^s$ into a sequence $\mathcal{S}_{m^s}$ that summarizes their semantics, which is then encoded with GRU to capture sequential dependencies.

\begin{equation}
    \boldsymbol{h}_{m^s}^i=\mathrm{GRU}_s(\boldsymbol{h}_{m^s}^{i-1}, \boldsymbol{e}_{m^s}[i])
    \label{equ: encode candidate sequence}
\end{equation}

\noindent where $\boldsymbol{e}_{m^s}[i]$ represents the embedding of $\mathcal{S}_{m^s,i}$ and $\boldsymbol{h}_{m^s}^i$ denotes the corresponding hidden state. For simplicity, the last hidden is denoted as $\boldsymbol{h}_{m^s}$ and treated as the abstracted result.

\noindent\textbf{2) Recommendation via Dual Effects Reference}

Finally, MSAM performs the recommendation by jointly referring to the abstracted historical prescriptions $\boldsymbol{o}$ and abstracted selected candidates $\boldsymbol{h}_{m^s}$:

\begin{equation}
    \hat{\boldsymbol{y}}=\sigma(f([\boldsymbol{q}^T, \boldsymbol{o}, \boldsymbol{h}_{m^s}]))
    \label{equ: stage II recommendation}
\end{equation}

\subsection{Training}

In Stage I, MSAM is trained by minimizing weighted loss $\mathcal{L}_I$ to jointly optimize recommendation performance and candidate selection, which contains a binary cross-entropy loss $\mathcal{L}_b$ for recommendation and $\mathcal{L}_{\alpha}$ that guides the model to select medications relevant to patient conditions. Here, $\alpha$ is the weight for $\mathcal{L}_{\alpha}$, $\boldsymbol{y}$ is the multi-hot vector corresponding to the ground truth $m^T$.

\begin{equation}
    \mathcal{L}_I=\mathcal{L}_b+\alpha\mathcal{L}_{\alpha}
    \label{equ: stage I loss}
\end{equation}

\begin{equation}
    \mathcal{L}_b=-\sum_{i=1}^{|S_m|}[\boldsymbol{y}_i \mathrm{log} \sigma(\hat{\boldsymbol{y}}_i)+(1-\boldsymbol{y}_i)\mathrm{log}(1-\sigma(\hat{\boldsymbol{y}}_i))]
    \label{equ: bce loss}
\end{equation}

\begin{equation}
    \mathcal{L}_{\alpha}=-\sum_{i=1}^{|S_m|}[\boldsymbol{y}_i \mathrm{log} \sigma(\boldsymbol{\alpha}_i)+(1-\boldsymbol{y}_i)\mathrm{log}(1-\sigma(\boldsymbol{\alpha}_i))]
    \label{equ: attention loss}
\end{equation}

In stage II, the candidate selection attention module is fixed and reused, and MSAM is optimized only with $\mathcal{L}_b$ to the refine recommendation performance.

\section{Experiments}

Experiments are designed to evaluate MSAM from four complementary perspectives: overall effectiveness, transferability to existing models, robustness of medication abstraction, and impact of semantic alignment.

\noindent\textbf{Q1:} Does MSAM outperform existing models across datasets and evaluation settings?

\noindent\textbf{Q2:} Can the semantic alignment strategy be integrated into existing methods to yield consistent improvements?

\noindent\textbf{Q3:} Is MSAM able to robustly abstract collective effects across varying sizes of medication sets?

\noindent\textbf{Q4:} How do different degrees of alignment between patient representations and references affect performance?

\begin{figure*}[b]
    \centering
    \includegraphics[width=0.95\linewidth]{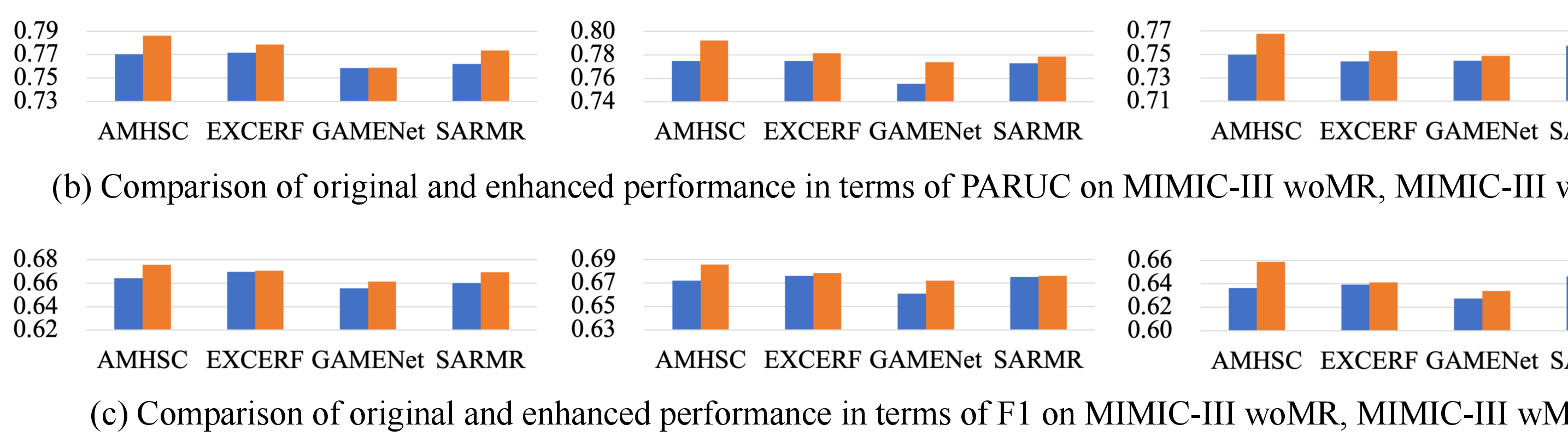}
    \caption{Comparison of four longitudinal-based models and their variants enhanced by abstracted medications.}
    \label{fig:comparison of original and enhanced performance}
\end{figure*}

\subsection{Experiment Settings}

\begin{table}[b]
    \centering
    \caption{Statistics of Selected Data}
    \begin{tabular}{lcccc}
    \hline
    {\makecell{ Datasets \\[-8pt] \rule{2.4cm}{0.4pt}}}     & \multicolumn{2}{c}{\makecell{ MIMIC-III \\[-8pt] \rule{2.2cm}{0.4pt}}}  & \multicolumn{2}{c}{\makecell{ MIMIC-IV \\[-8pt] \rule{2.2cm}{0.4pt}}} \\
    \multicolumn{1}{c}{Restrictions} & woMR    & wMR & woMR  & wMR  \\ \hline
    \# patients         & 6350  & 6350  & 8522  & 8522 \\
    \# diagnosis        & 1958  & 1958  & 1962  & 1962 \\
    \# procedure        & 1430  & 1430  & 1338  & 1338 \\ 
    \# medication       & 151   & 131   & 151   & 130  \\
    avg\# of visit      & 2.37  & 2.37  & 2.26  & 2.26 \\
    avg\# of diagnosis  & 13.64 & 13.64 & 17.56 & 17.56\\
    avg\# of procedure  & 4.54  & 4.54  & 3.70  & 3.70 \\
    avg\# of medication & 19.86 & 19.20 & 17.79 & 16.85 \\
    \hline
    \end{tabular}
    \label{tab:dataset statistics}
\end{table}

\noindent\textbf{Datasets:} We evaluate MSAM on two real-world clinical datasets MIMIC-III and MIMIC-IV \cite{johnson2016mimic,johnson2023mimic}. Patients with at least two visits are selected for experiments, and their diagnoses and procedures are used as inputs for generating corresponding medications. To conduct a fair comparison with baselines that incorporate drug molecular structures, we filter medications to those with well-defined molecular representations, and summarize results under two conditions: without molecular restrictions (woMR) and with molecular restrictions (wMR). Detailed statistics under different settings are provided in Table~\ref{tab:dataset statistics}.

\noindent\textbf{Baselines:} We compare MSAM with different baseline methods that incorporate patient records for medication recommendation in various forms, including AMHSC \cite{wang2022adaptive}, EXCERF \cite{wang2025beyond}, GAMENet\cite{shang2019gamenet}, Leap \cite{zhang2017leap}, RETAIN \cite{choi2016retain}, SARMR \cite{wang2021self} and VITA \cite{kim2024vita}. We also compare MSAM with models that rely on drug molecular structures, namely SafeDrug \cite{yang2021safedrug} and MoleRec \cite{yang2023molerec}. All baselines are evaluated in terms of three widely adopted metrics: Jaccard Similarity, Precision Recall AUC (PRAUC), and F1.

\subsection{\textbf{Q1:} Comparison with Baseline Methods}

Table \ref{tab:comparison with baselines on mimic-iii} and Table \ref{tab:comparison with baselines on mimic-iv} summarize the performance of MSAM and baseline methods under different settings. Since SafeDrug and MoleRec depend on molecular structures for recommendation, they are not evaluated under the woMR setting. As both tables show, MSAM consistently achieves the best performance in terms of all metrics, which indicates that MSAM uncovers and abstracts the multi-level structures inherent in medications into references, enabling the model to obtain their collective therapeutic effects that match with high-level patient conditions. By bridging the clinical semantic gap, MSAM achieves substantial gains over all competing methods. Meanwhile, several baselines retrieve references by mean-pooling or independent attention over individual drugs, thereby failing to capture multi-level structures within medications. Consequently, these methods overlook the semantic mismatch between referenced medications and patient conditions and fail to compete with MSAM. Regarding methods that rely solely on patient records, the absence of references limits their ability to form a robust basis for the recommendation and leads to inferior performance. Moreover, MSAM also surpasses models that incorporate drug molecular structures. In summary, the results address \textbf{Q1:} MSAM achieve superior performance over state-of-the-art baselines.

\begin{table}[t]
    \centering
    \caption{Comparison of Different Methods on MIMIC-III}
    \begin{tabular}{cccc}
    \hline
    {\makecell{ Datasets \\[-8pt] \rule{1.5cm}{0.4pt}}}     & \multicolumn{3}{c}{\makecell{ woMR/wMR \\[-8pt] \rule{6cm}{0.4pt}}}  \\
    Measures & Jaccard    & PRAUC & F1   \\ 
    \hline
    AMHSC    & 0.508/0.516 & 0.770/0.775 & 0.664/0.672 \\
    EXCERF   & 0.514/0.521 & 0.772/0.775 & 0.670/0.676 \\
    GAMENet  & 0.499/0.505 & 0.759/0.755 & 0.656/0.661 \\
    Leap     & 0.448/0.421 & 0.647/0.587 & 0.611/0.584 \\
    RETAIN   & 0.488/0.490 & 0.750/0.755 & 0.648/0.650 \\
    SARMR    & 0.504/0.520 & 0.762/0.773 & 0.660/0.675 \\
    VITA     & 0.524/0.524 & 0.762/0.760 & 0.678/0.678 \\
    SafeDrug &\parbox[c][0cm][c]{0.8cm}{\centering\rule{0.5cm}{0.4pt}}/0.513 & \parbox[c][0cm][c]{0.8cm}{\centering\rule{0.5cm}{0.4pt}}/0.766 & \parbox[c][0cm][c]{0.8cm}{\centering\rule{0.5cm}{0.4pt}}/0.669 \\
    MoleRec  &\parbox[c][0cm][c]{0.8cm}{\centering\rule{0.5cm}{0.4pt}}/0.526 & \parbox[c][0cm][c]{0.8cm}{\centering\rule{0.5cm}{0.4pt}}/0.769 & \parbox[c][0cm][c]{0.8cm}{\centering\rule{0.5cm}{0.4pt}}/0.680 \\
    MSAM     & 0.527/0.532 & 0.780/0.787 & 0.682/0.686 \\
    \hline
    \end{tabular}
    \label{tab:comparison with baselines on mimic-iii}
\end{table}

\begin{table}[t]
    \centering
    \caption{Comparison of Different Methods on MIMIC-IV}
    \begin{tabular}{cccc}
    \hline
    {\makecell{ Datasets \\[-8pt] \rule{1.5cm}{0.4pt}}}     & \multicolumn{3}{c}{\makecell{ woMR/wMR \\[-8pt] \rule{6cm}{0.4pt}}}  \\
    Measures & Jaccard    & PRAUC & F1   \\ 
    \hline
    AMHSC    & 0.479/0.489 & 0.750/0.754 & 0.636/0.646 \\
    EXCERF   & 0.482/0.481 & 0.744/0.748 & 0.639/0.639 \\
    GAMENet  & 0.469/0.474 & 0.745/0.744 & 0.627/0.632 \\
    Leap     & 0.426/0.439 & 0.623/0.624 & 0.588/0.600 \\
    RETAIN   & 0.455/0.457 & 0.725/0.724 & 0.615/0.617 \\
    SARMR    & 0.490/0.488 & 0.758/0.756 & 0.646/0.645 \\
    VITA     & 0.486/0.483 & 0.725/0.730 & 0.642/0.640 \\
    SafeDrug &\parbox[c][0cm][c]{0.8cm}{\centering\rule{0.5cm}{0.4pt}}/0.482 &\parbox[c][0cm][c]{0.8cm}{\centering\rule{0.5cm}{0.4pt}}/0.725 &\parbox[c][0cm][c]{0.8cm}{\centering\rule{0.5cm}{0.4pt}}/0.634 \\
    MoleRec &\parbox[c][0cm][c]{0.8cm}{\centering\rule{0.5cm}{0.4pt}}/0.489 &\parbox[c][0cm][c]{0.8cm}{\centering\rule{0.5cm}{0.4pt}}/0.732 &\parbox[c][0cm][c]{0.8cm}{\centering\rule{0.5cm}{0.4pt}}/0.647 \\
    MSAM     & 0.502/0.503 & 0.762/0.762 & 0.657/0.659 \\
    \hline
    \end{tabular}
    \label{tab:comparison with baselines on mimic-iv}
\end{table}

\subsection{\textbf{Q2:} Compatibility to Existing Methods}

To verify the compatibility of the medication abstraction strategy, we select longitudinal-based baseline methods, including AMHSC, EXCERF, GAMENet and SARMR, which treat medications as individuals, and replace their medication reference modules with the abstraction module. The comparison between the original models and their variants is shown in Figure \ref{fig:comparison of original and enhanced performance}. As illustrated, all variants consistently outperform the original methods across all datasets, highlighting the effectiveness of aligning patient conditions with referenced medications for recommendation. Notably, the improvements in PRAUC across all models indicate that medication abstraction derives the overall therapeutic effects of medications, which enhances the ranking quality of predictions when integrated with patient conditions. It is worth noting that EXCERF exhibits relatively smaller gains compared to others. This is likely due to the incorporation of structured clinical knowledge from diagnoses and procedures, which leads to semantically rich patient representations that are harder for the abstracted medications to align with. This observation suggests an interesting direction for future work: adaptively determining the semantic level at which patient conditions should integrate with referenced medications instead of relying on fixed matching layer. In summary, the results address \textbf{Q2:} semantic gap bridging is broadly applicable to improve the recommendation performance of existing methods.

\begin{figure}[t]
    \centering
    \includegraphics[width=1\linewidth]{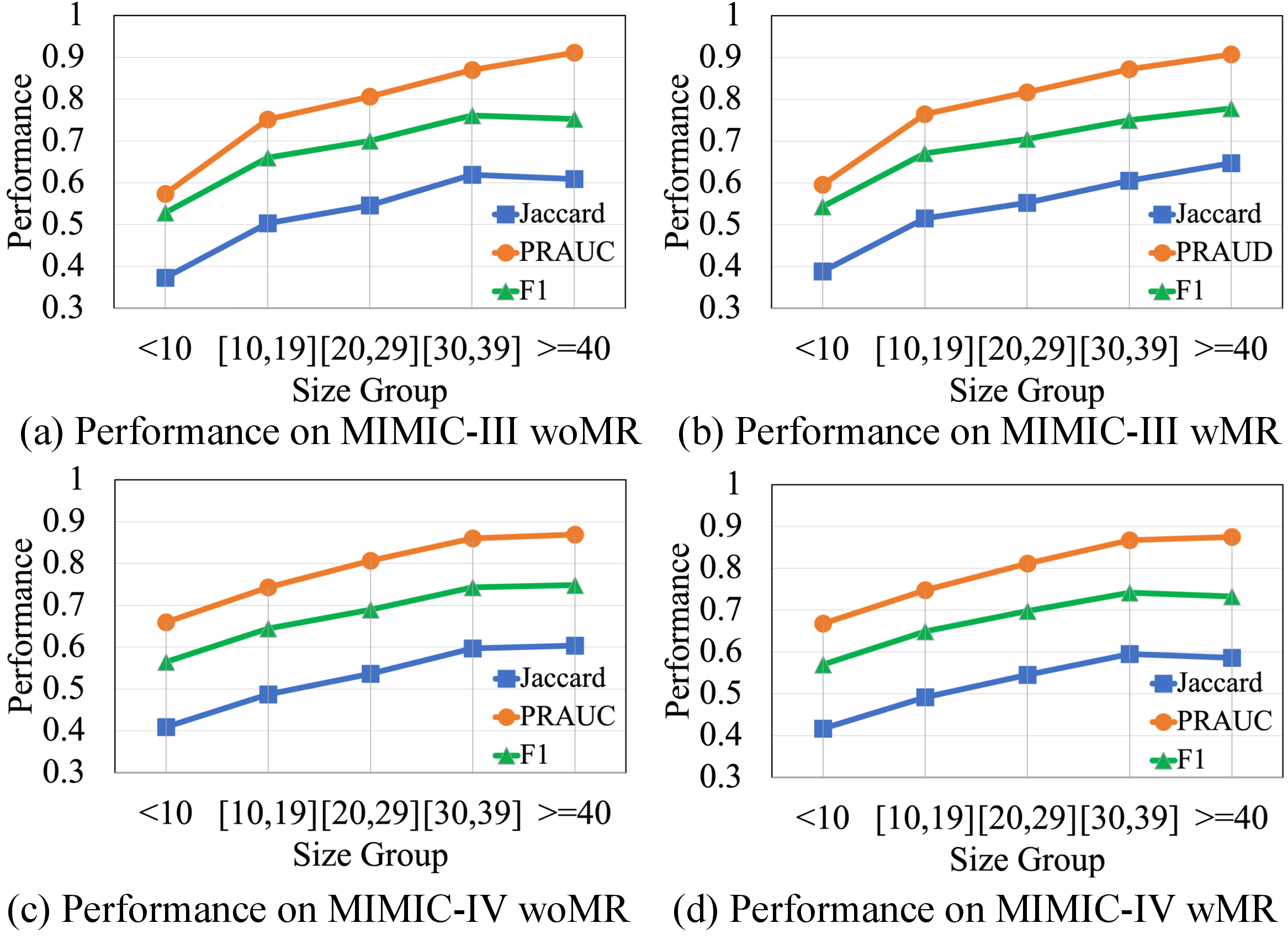}
    \caption{Performance in Varying Size of Medication Sets}
    \label{fig:robustness}
\end{figure}

\subsection{\textbf{Q3:} Robustness on Size of Medication Sets}

To verify the robustness of MSAM under varying medication set sizes, we evaluate its performance on patient cohorts with different average numbers of medications in their historical prescriptions. As illustrated in Figure \ref{fig:robustness}, MSAM achieves consistent trends on MIMIC-III and MIMIC-IV under the woMR and wMR settings. Specifically, performance improves as the number of medications increases and approaches or exceeds the overall performance of MSAM when the average number of medications exceeds 10, suggesting that small medications fail to provide sufficient clinical semantics as informative references. When the average number of medications exceeds 40, performance saturates or shows slight decline, which is likely attributed to the limited capacity of RNN to model very long sequences. Overall, the results answer \textbf{Q3:} MSAM generates stable abstractions of collective effects across varying sizes of medication sets, indicating robustness of the underlying medication ordering mechanism.

\subsection{\textbf{Q4:} Ablation Study}

To examine the necessity of each component and how different degrees of alignment affect the performance, we compare MSAM with the following ablation variants:
(1) $\textrm{MSAM-V}_{\text{set}}$ averages event embeddings within each visit as conditions without longitudinal modeling across visits; (2) $\textrm{MSAM-M}_{\text{set}}$ treats historical prescriptions and candidate medications as sets without leveraging inherent semantics; (3) $\textrm{MSAM-P}_{\text{set}}$ treats historical prescriptions as sets while leveraging semantics in the selected candidate medications; (4) $\textrm{MSAM-C}_{\text{att}}$ preserves multi-level structures in historical prescriptions but refers to candidate medications as sets.

The results under woMR and wMR settings for MIMIC-III and MIMIC-IV are summarized in Table \ref{tab:ablation study on mimic-iii} and Table \ref{tab:ablation study on mimic-iv} respectively. As shown, MSAM surpasses all variants across metrics, indicating the importance of semantic alignment between patient conditions and referenced medications and the necessity of each component. Specifically, $\textrm{MSAM-V}_{\text{set}}$ and $\textrm{MSAM-M}_{\text{set}}$ face similar limitations: the semantic mismatch between patient representations and medication references prevent them from competing with MSAM. Meanwhile, $\textrm{MSAM-P}_{\text{set}}$ and $\textrm{MSAM-C}_{\text{att}}$ capture semantics only from one of the two reference sources. The partial utilization and mismatch between the two sources lead to performance degradation. Notably, $\textrm{MSAM-P}_{\text{set}}$ shows a substantial decline in PRAUC, likely due to the overlook of therapeutic patterns from historical prescriptions and the semantic mismatch with candidate medications. In comparison, $\textrm{MSAM-M}_{\text{set}}$ treats both references as sets to avoid such mismatch, while $\textrm{MSAM-C}_{\text{att}}$ preserves  the multi-level structures within historical prescriptions. In summary, the results address \textbf{Q4:} effective recommendation relies on the semantic alignment between (i) patient conditions and referenced medications; (ii) the two sources of references. Together, they indicate that the observed performance gains stem from uncovering and aligning collective effects with patient conditions.

\begin{table}[t]
    \centering
    \caption{Ablation Study Results on MIMIC-III}
    \begin{tabular}{cccc}
    \hline
    {\makecell{ Datasets \\[-8pt] \rule{1.5cm}{0.4pt}}}     & \multicolumn{3}{c}{\makecell{ woMR/wMR \\[-8pt] \rule{6cm}{0.4pt}}}  \\
    Measures & Jaccard    & PRAUC & F1   \\ 
    \hline
    $\textrm{MSAM-V}_{\text{set}}$  & 0.519/0.526 & 0.771/0.779 & 0.674/0.680 \\
    $\textrm{MSAM-M}_{\text{set}}$  & 0.519/0.523 & 0.775/0.782 & 0.674/0.677 \\
    $\textrm{MSAM-P}_{\text{set}}$ & 0.522/0.528 & 0.666/0.668 & 0.677/0.683 \\
    $\textrm{MSAM-C}_{\text{att}}$ & 0.518/0.519 & 0.774/0.778 & 0.673/0.674 \\
    MSAM  & 0.527/0.532 & 0.780/0.787 & 0.682/0.686 \\
    \hline
    \end{tabular}
    \label{tab:ablation study on mimic-iii}
\end{table}

\begin{table}[t]
    \centering
    \caption{Ablation Study Results on MIMIC-IV}
    \begin{tabular}{cccc}
    \hline
    {\makecell{ Datasets \\[-8pt] \rule{1.5cm}{0.4pt}}}     & \multicolumn{3}{c}{\makecell{ woMR/wMR \\[-8pt] \rule{6cm}{0.4pt}}}  \\
    Measures & Jaccard    & PRAUC & F1   \\ 
    \hline
    $\textrm{MSAM-V}_{\text{set}}$  & 0.495/0.497 & 0.753/0.754 & 0.651/0.653 \\
    $\textrm{MSAM-M}_{\text{set}}$  & 0.495/0.494 & 0.755/0.756 & 0.650/0.650 \\
    $\textrm{MSAM-P}_{\text{set}}$ & 0.499/0.497 & 0.639/0.636 & 0.654/0.653 \\
    $\textrm{MSAM-C}_{\text{att}}$ & 0.495/0.496 & 0.756/0.755 & 0.650/0.652 \\
    MSAM  & 0.502/0.503 & 0.762/0.762 & 0.657/0.659 \\
    \hline
    \end{tabular}
    \label{tab:ablation study on mimic-iv}
\end{table}

\section{Conclusion}

In this paper, we propose MSAM, a novel medication recommendation model that bridges the clinical semantic gap between patient conditions and referenced medications. Unlike existing methods that leverage high-level patient representations describing complex conditions as a query to refer individual drugs, MSAME perform multi-level medication abstraction to capture collective therapeutic effects aligned with patient conditions as informative references. The model introduces a multi-head graph reasoning mechanism to organize flat daily medication sets into clinically meaningful semantic units in the absence of intrinsic composition rules, enabling feature propagation from individual drugs to higher-level representations and supporting two-stage abstraction over historical prescriptions and candidate medications across heterogeneous clinical structures. Experiments on two widely used real-world clinical datasets show that MSAM consistently outperforms baseline methods across multiple metrics, remains robust to varying medication set sizes, and improves performance of existing recommendation methods when integrated with them, indicating its ability to model and align collective medication effects with patient conditions for improved recommendation.

\appendix




\bibliographystyle{named}
\bibliography{ijcai26}

\end{document}